\documentclass{article}
\usepackage[utf8]{inputenc}
\usepackage{authblk}
\usepackage{graphicx}
\graphicspath{{noiseimages/}}
\usepackage{xcolor}
\usepackage[T2A]{fontenc}
\usepackage{amssymb}
\usepackage{tikz}
\usepackage{pgfplots}
\usepackage{mathrsfs}
\pgfplotsset{compat=1.7}
\usetikzlibrary{intersections, pgfplots.fillbetween}
\usetikzlibrary{decorations.pathmorphing}
\usetikzlibrary{arrows.meta}
\usepackage[mode=buildnew]{standalone}
\usepackage[toc,page]{appendix}
\usepackage{amsmath}
\usepackage{bm}
\usepackage{float}
\usepackage{comment}
\usepackage{a4wide}
\usepackage[english]{babel}
\usepackage{hyperref}

\definecolor{urlcolor}{HTML}{990000}
\definecolor{linkcolor}{HTML}{005F5F}
\hypersetup{
    pdfstartview=FitH,
    linkcolor=linkcolor,
    urlcolor=urlcolor,
    colorlinks=true,
    citecolor=blue
}

\author[1,2]{D. V. Diakonov\footnote{\tt dmitrii.dyakonov@phystech.edu}}

\affil[1]{Institutskii per. 9, Moscow Institute of Physics and Technology,
141700, Dolgoprudny, Russia}
\affil[2]{\itshape Bol'shoi Karetnyi per., 19, Institute for Information
Transmission Problems, 127994, Moscow, Russia}

\title{\textcolor{black}{Soliton like solutions of $\lambda \phi^4$ theory in de Sitter space}}
\date{}

\begin{document}
\maketitle

\begin{abstract}
We study classical solutions of a scalar field with quartic potential in de Sitter space-time. Using the ansatz $\phi=F(X\cdot\xi)$, via embeding coordinate $X_\alpha$ of the ambiend space we find several families of analytic solutions classified by the causal character of the constant vector $\xi$. For time-like $\xi$, the solutions are globally regular and describe either global oscillations around a minimum or global vacuum transition; for space-like $\xi$, they are singular in the global patch but regular in the open slicing. In the Poincaré patch they correspond to contracting bubbles. We compute the classical actions of these solition which are finite only in spacetime dimensions $D \le 5$. We show that the $D=4$ global vacuum transition solution has the radiation equation of state $w=1/3$. In $D=4$ global vacuum transition solution, the retarded Green's function acquires a growing tail as compared to the Green's function on top of vacuum, implying that the transition amplifies the signal at late times.
\end{abstract}

\newpage

\section{Introduction}

Domain walls are a common prediction of particle-physics models in which a
discrete symmetry is spontaneously broken. Whenever the vacuum manifold of such a
theory consists of several disconnected components, regions of the Universe that
have never been in causal contact may be left in different vacua, and the field
configurations interpolating between them are precisely the walls. Their
cosmological relevance is well known \cite{Zeldovich:1974uw, Kibble:1976sj}: the
energy density stored in a wall network redshifts much more slowly than that of
matter or radiation, so that even a modest population of walls tends to come to
dominate the expansion. The requirement that this does not happen turns the
question of whether walls form at all into a sharp constraint on the underlying
model.

Apart from that, quantum field theory beyond tree level in de Sitter space is
very challenging, and initial conditions dramatically change the non-peretrubative behavior of the
field theory due to secular effects
\cite{Krotov:2010ma,Akhmedov:2013vka,Akhmedov:2024npw,Akhmedov:2019cfd,Akhmedov:2022uug,Akhmedov:2021rhq,Miao:2024shs,Moreau:2020gib,Moreau:2019jpn}.
Therefore, considering quantum field theory on nontrivial classical backgrounds
may shed light on the behavior of quantum fields in de Sitter space.

In this paper we consider scalar field theory with a quartic potential and obtain
several families of analytic solutions in different dimensions. We treat both
signs of $m^2$ on the same footing: for $m^2>0$ the potential has a single
minimum, and the solutions simply describe the field oscillating around it,
whereas for $m^2<0$ the potential is of the double-well form with two
degenerate vacuua, and the corresponding solutions interpolate between them,
i.e.\ they are of the global domain-wall type. The solutions that we consider do not deform the background de Sitter metric.

The solutions fall into two classes according to the causal character of the
vector $\xi$. For time-like $\xi$ we find four families, all of which are
globally well defined: in the global slicing the field starts in a vacuum at past
infinity and ends, depending on the case, either in the same vacuum or in the
other one. For space-like $\xi$ the configurations are singular on some
hypersurface of global de Sitter space, but they are perfectly regular in other
patches -- in particular in the open slicing -- and are of interest for that
reason. The space-like hypersurface in the
Poincaré patch looks like a contracting ball. For each regular solution we also compute the classical action; it turns
out to be finite only in relatively low dimensions. One noteworthy result is that
in $3+1$ dimensions the equation of state of one of our solutions is $w=1/3$,
i.e.\ exactly the same as for radiation.

\section{Geometry}
\label{sec:geometry}

A maximally symmetric $D$-dimensional space-time can be embedded in a flat
ambient space-time of dimension $D+1$. In particular, de Sitter
spacetime is the hyperboloid embedded in a $(D+1)$-dimensional ambient flat
space-time with signature $(-,+,\ldots,+)$:
\begin{align}
dS_D
=
\left\{
X \in \mathbb{R}^{1,D}
:\;
X\cdot X = X_\alpha X^\alpha = R^2
\right\},
\qquad \alpha = 0,1,\ldots,D .
\end{align}
Throughout the paper we set the de Sitter radius to $R=1$.

Let us define the function
\begin{align}
f(X) = X\cdot \xi,
\end{align}
where $\xi$ is a constant vector in the ambient space-time. This function
satisfies the Klein--Gordon equation
\begin{align}
\label{box12}
\Box (X\cdot \xi)
=
-D (X\cdot \xi),
\end{align}
and obeys the identity
\begin{align}
\nabla_\mu (X\cdot \xi_1 ) \nabla^\mu (X \cdot \xi_2)
=
- (X\cdot \xi_1 ) (X \cdot \xi_2) + (\xi_1 \cdot \xi_2).
\label{waveprop}
\end{align}
Both relations follow from the fact that the covariant derivative on the
hyperboloid is the ambient derivative projected onto the tangent space. For more
details, see \cite{Akhmedov:2026msi}. This property holds for any maximally
symmetric space
\cite{Bros:1994dn,Bros:1995js,Moschella:2025lqy,Moschella:2007zza,Akhmedova:2019bau}.

For a function $F[(X\cdot\xi)]$, this implies that
\begin{align}
\label{reduction}
\Box F[(X\cdot\xi)] =\bigl[(\xi\cdot\xi)-(X\cdot\xi)^{2}\bigr]F''[(X\cdot\xi)]
-D (X\cdot\xi) F'[(X\cdot\xi)],
\end{align}
i.e.\ the ansatz $\phi=F(X\cdot\xi)$ reduces equation of motion of the form
$\Box\phi=V'(\phi)$ to an ordinary differential equation in a single variable.
Using this approach, one can find solutions in different models
\cite{Akhmedov:2026msi,Sadekov:2026gmc,Diakonov:2026cbc}.

De Sitter space can be covered by several coordinate systems. For example, in
global coordinates
\[
{\displaystyle
\begin{aligned}
X^0&=R\sinh (t),\\
X^{i}&=R\cosh (t) \Omega^{i},\qquad
\Omega\in S^{D-1},
\end{aligned}}
\]
with the metric
\[
{\displaystyle
ds^{2}=-dt^{2}+\cosh^{2}(t) d\Omega_{D-1}^{2}.
}
\]
One can also use the conformal time $T$, defined by
\begin{align}
\cos (T)=\frac{1}{\cosh (t)},\qquad \tan (T)=\sinh (t),\qquad
T\in\Bigl(-\frac{\pi}{2},\frac{\pi}{2}\Bigr).
\end{align}

For a part of the solutions that we present below $(\xi\cdot\xi)<0$, i.e.\ $\xi$
is time-like, so that an $SO(1,D)$ rotation brings it to the form
\begin{align}
\xi=|\xi| (-1,\bm{0}),\qquad |\xi|=\sqrt{|(\xi\cdot\xi)|}.
\end{align}
Hence
\begin{align}
(X\cdot\xi)=|\xi|\sinh (t)=|\xi|\tan (T) .
\end{align}

Another example is the open slicing:
\[
{\displaystyle
\begin{aligned}
X_{0}&=\sinh(t)\cosh (\chi),\\
X_{1}&=\cosh(t),\\
X_{i}&=z_{i}\sinh(t)\sinh (\chi),\qquad 2\leq i\leq D,
\end{aligned}}
\]
with the metric
\[
{\displaystyle
ds^{2}=-dt^{2}+\sinh^{2}(t) dH_{D-1}^{2},
}
\]
where $dH_{D-1}^2$ is the metric of the unit hyperbolic space $H_{D-1}$. For the
remaining solutions presented below $(\xi\cdot\xi)>0$, i.e.\ $\xi$ is
space-like; a rotation then aligns it with the $X^{1}$ axis, so that
\begin{align}
    (X\cdot\xi)= |\xi| \cosh(t),\qquad |\xi|=\sqrt{(\xi\cdot\xi)} .
\end{align}

\section{Classical solutions}
\label{sec:solutions}

We consider the action of a scalar field theory with a potential:
\begin{align}
\label{action}
S= -\int d^D x \sqrt{g} \left( \frac{1}{2} \nabla_\mu \phi \nabla^\mu \phi
+V(\phi)\right),\qquad
V(\phi)=\frac{1}{2}m^2\phi^2+\frac{\lambda}{4} \phi^4 .
\end{align}
The equation of motion is
\begin{align}
\label{eom}
\Box \phi -m^2 \phi -\lambda \phi ^3=0.
\end{align}
We consider both signs of $m^{2}$: for $m^{2}>0$ the potential has a single
minimum at $\phi=0$, whereas for $m^{2}<0$ it is of the double-well
form, with two degenerate minima at $\phi_{v}=\pm\sqrt{-m^{2}/\lambda}$. In the
latter case we should normalize the potential so that it vanishes in the vacuum:
\begin{align}
\label{Vsub}
V(\phi)=\frac{\lambda}{4}\bigl(\phi^{2}-\phi_{v}^{2}\bigr)^{2}
\end{align}
which differs from \eqref{action} by the constant $m^{4}/4\lambda$. This constant does
not affect the equation of motion \eqref{eom}, but it is essential below: without it the
classical action of any configuration approaching a vacuum at infinity would diverge with
the volume of de Sitter space, and the stress-energy tensor would acquire a cosmological like terms $\tfrac{m^{4}}{4\lambda} \delta^{\mu}_{\nu}$.

\subsection{Solutions with a time-like vector $\xi$}

We find four classes of solutions with a time-like vector $\xi$, which are
presented in Table~\ref{tab:Sol}. These solutions are globally well defined and
describe how the field evolves from a vacuum value $\phi_v$ at past infinity either to the
other vacuum or back to the same one. To give a simple explanation why the field evolves in this
way, let us consider the equation of motion in global coordinates:
\begin{align}
-\partial_t^2\phi-(D-1)\tanh t \partial_t\phi
+
\frac{1}{\cosh^2 t} \Delta_{S^{D-1}}\phi -m^2 \phi -\lambda \phi ^3=0,
\end{align}
where the second term acts like friction: for $t < 0$, this friction is
negative---precisely what causes the field to grow---whereas for $t > 0$, the
sign becomes positive, causing the solution to fall back into the minimum.

\begin{table}[h]
\centering
\footnotesize
\caption{Solutions with a time-like vector $\xi$.}
\label{tab:Sol}
\setlength{\tabcolsep}{3pt}
\renewcommand{\arraystretch}{1.8}
\begin{tabular}{|c|l|l|l|l|l|l|}
\hline
No. & $\phi$ & $m^2$ & $\xi\cdot\xi$ & a& $D$ & $\phi_v$\\
\hline
1 & $\frac{b}{\sqrt{1+(X\cdot\xi)^2}}$ & $0$ & $-1$ &
$b=\dfrac{1}{\sqrt{\lambda}}$ & $D=2$ & $0$\\
\hline
2 & $ \frac{b (X\cdot\xi)}{\sqrt{1+(X\cdot\xi)^2}}$& $3-D$ & $-1$ &
$b=\phi_v $& $D\ge4$ &
$\pm\sqrt{\dfrac{D-3}{\lambda}}$\\
\hline
3 & $\frac{b(X\cdot\xi)}{1+(X\cdot\xi)^2}$ & $D-2$ & $-\dfrac{D-1}{3}$ &
$b=\dfrac{2\sqrt{2/3}\sqrt{4-D}}{\sqrt{\lambda}}$ & $D\le3$ & $0$ \\
\hline
$4^{\pm}$& $\phi_v+\frac{b}{1+(X\cdot\xi)^2}$& $3-D$ &
$\pm\dfrac{\sqrt{(D-1)(D-3)}}{\sqrt{3}}-\frac{2(D-1)}{3}$ &
\shortstack{
$b=\dfrac{ 2(\pm\sqrt{3(D-1)}-3\sqrt{D-3})}{3\sqrt{\lambda}}$}& $D\ge3$ &
$\pm\sqrt{\dfrac{D-3}{\lambda}}$ \\
\hline
\end{tabular}
\end{table}

For cases 1, 3 and $4^\pm$ with $D=3$, the square of the mass is
non-negative, i.e.\ the vacuum is non-degenerate. Hence, these solutions describe
just oscillations of the field: at past and future infinity of the global
coordinate system the field sits at its vacuum value.

\begin{figure}[H]
    \centering
    \includegraphics[scale=0.6]{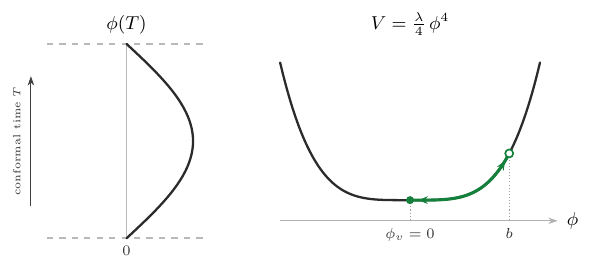}
    \caption{Case 1 for $D=2$. The field starts and ends at the
    non-degenerate vacuum $\phi_v=0$.}
    \label{pic1}
\end{figure}

For case 1, which exists only for $D=2$, with $m^2=0$ and $\xi=|\xi| (-1,\bm{0})$, we have
\begin{align}
\phi=\frac{1}{\sqrt{\lambda}}\cos T,
\end{align}
which is shown in Figure~\ref{pic1}.

\begin{figure}[H]
    \centering
    \includegraphics[scale=0.6]{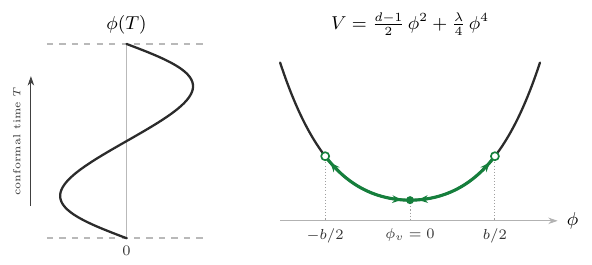}
    \caption{Case 3. The field
    returns to the vacuum $\phi_v=0$ at both past and future infinity.}
    \label{pic2}
\end{figure}

For case 3, which exists for $D\le3$, the general solution is
\begin{align}
\phi=\frac{b(X\cdot\xi)}{1+(X\cdot\xi)^2},
\qquad
m^2=D-2,\quad
\xi\cdot\xi=-\frac{D-1}{3},\quad
b=\frac{2\sqrt{2/3}\sqrt{4-D}}{\sqrt{\lambda}}.
\end{align}
If $\xi=|\xi| (-1,\bm{0})$, then:
\begin{align}
\phi=\frac{b |\xi| \sin T\cos T}{\cos^{2}T+|\xi|^{2}\sin^{2}T},
\end{align}
which is shown in Figure~\ref{pic2}.

\begin{figure}[H]
    \centering
    \includegraphics[scale=0.6]{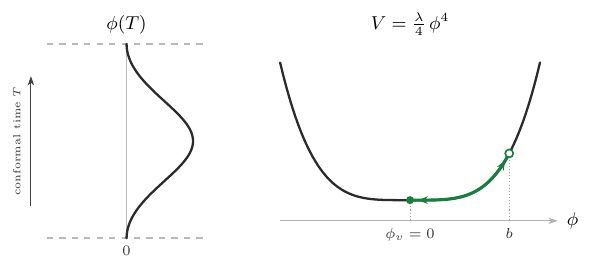}
    \caption{Case 4 for $D=3$, where the two branches coincide,
    $4^+=4^-$. The solution is globally regular and starts and ends at the
    vacuum $\phi_v=0$.}
    \label{pic3}
\end{figure}

For case 4 with $D=3$, the two branches coincide, $4^+=4^-$. In this case $\xi=|\xi| (-1,\bm{0})$,
$\phi_v=0$, $m^2=0$, $(\xi\cdot\xi)=-4/3$, and
\begin{align}
\phi
=
\pm\frac{2}{3}\sqrt{\frac{6}{\lambda}}
\frac{\cos^{2}T}{\cos^{2}T+\frac{4}{3}\sin^{2}T},
\end{align}
which is shown in Figure~\ref{pic3}.

For solutions 2 and $4^\pm$ with $D \ge 4$, the vacuum is degenerate, because we
have the double-well potential. Solution 2 describes how the field transitions from
one vacuum value to another.

\begin{figure}[H]
    \centering
    \includegraphics[scale=0.6]{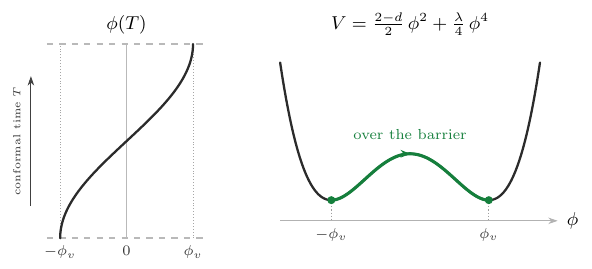}
    \caption{Case 2 : the time-like global vacuum transition between the degenerate vacua.}
    \label{pic4}
\end{figure}

For case 2, the general solution is
\begin{align}
\phi=\frac{b (X\cdot\xi)}{\sqrt{1+(X\cdot\xi)^2}},
\qquad
m^2=3-D,\quad
\xi\cdot\xi=-1,\quad
b=\phi_v=\pm\sqrt{\frac{D-3}{\lambda}},
\end{align}
valid for $D\ge4$. When ($\xi=|\xi| (-1,\bm{0})$,
we have
\begin{align}
\phi=\sqrt{\frac{D-3}{\lambda}}\sin T,
\end{align}
which is shown in Figure~\ref{pic4}.

\begin{figure}[H]
    \centering
    \includegraphics[scale=0.6]{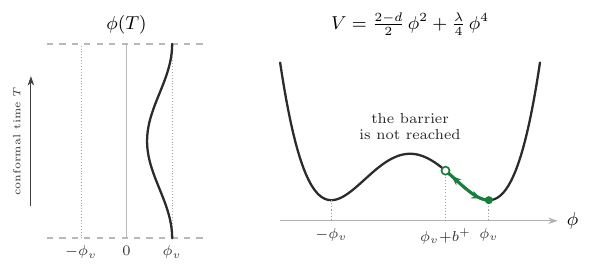}
    \caption{Typical branch $4^+$ for $D>4$. The field starts in one of the
    degenerate vacua,, and returns to the same vacuum state.}
    \label{pic5}
\end{figure}

\begin{figure}[H]
    \centering
    \includegraphics[scale=0.6]{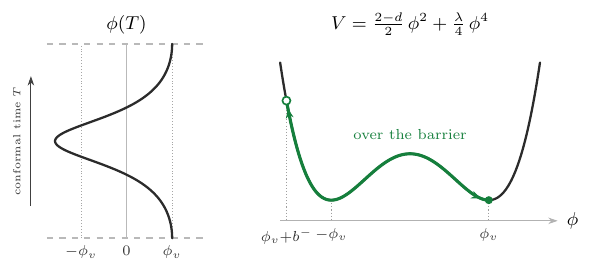}
    \caption{Typical branch $4^-$ for $D\ge 4$. The field starts in one of the
    degenerate vacua, and returns to the same
    vacuum state.}
    \label{pic6}
\end{figure}

Similarly, for case 4, the solution is arranged in such a way that the field is in the
vacuum in the past and then, depending on the branch of the solution, either
oscillates slightly or quite strongly, but eventually returns to the same vacuum
value. Two typical cases are shown in Figures~\ref{pic5} and~\ref{pic6}. Note
that for $D=4$ the branch $4^+$ degenerates into the constant vacuum
configuration, $b=0$, so that only the branch $4^-$ is non-trivial.

Finally, for each solution we can compute the classical action. It turns out that
all the solutions have a finite action only in relatively low dimensions. More
details are given in Table~\ref{tab:Scl}. It is worth noting that they are solution in the
Lorentzian signature. That meant that if we want to compute path inegral in this theory we should take into account all classical solutions with finite action:
\begin{align}
   Z= \sum_{cl.\ sol.} \int \mathcal{D} \phi e^{i S[\phi_{cl}+\phi]}.
\end{align}

\begin{table}[H]
\centering
\caption{Classical action $S_{\rm cl}$ for the solutions of Table~\ref{tab:Sol}.}
\label{tab:Scl}
\begin{tabular}{|c|c|l|}
\hline
Case & $D$ & $S_{\rm cl}$ \\
\hline
1 & 2 & $\displaystyle \frac{\pi^2}{4\lambda}$ \\
\hline
2 & 4 & $\displaystyle \frac{\pi^3}{2\lambda}$ \\
2 & 5 & $\displaystyle \frac{8\pi^2}{3\lambda}$ \\
2 & $\ge 6$ & diverges; in dimensional regularization:
$-\dfrac{(D-5)(D-3) \pi^{(D+1)/2}
\Gamma \left(\frac{5}{2}-\frac{D}{2}\right)}{2 \lambda
\Gamma \left(3-\frac{D}{2}\right)
\Gamma \left(\frac{D}{2}\right)}$\\
\hline
3 & 2 & $\displaystyle \frac{8 \pi ^2}{3 \sqrt{3}\lambda}$\\
3 & 3 & $\displaystyle \frac{32 \pi  \tanh
^{-1}\left(\frac{1}{\sqrt{3}}\right)}{
3 \sqrt{3} \lambda}-\frac{16 \pi }{\lambda}$\\
\hline
$4^+=4^-$& 3 & $\displaystyle \frac{32 \pi^2}{9\sqrt{3}\lambda} $\\
$4^+$& 4& $0$ \\
$4^-$& 4& $\displaystyle \frac{16 \pi ^3}{\sqrt{3} \lambda}$\\
$4^\pm$& $\ge 5$ & diverges; in dimensional regularization, pole $1/\varepsilon$ \\
\hline
\end{tabular}
\end{table}

\subsection{Solutions with a space-like vector $\xi$}

Let us also note that there is another set of solutions which diverge at some
points of global de Sitter space. They are nevertheless of interest, since they
are regular in other patches -- for example, in the open slicing. These solutions
are collected in Table~\ref{tab:Sol2}. Note also that we could generalize cases
1 and 2 from Table~\ref{tab:Sol} with $\xi^2>0$ but they are not regular in any coordinate
system. Hence in Table~\ref{tab:Sol2} we present only solution that can be regular
in some coordinate system.

\begin{table}[h]
\centering
\footnotesize
\caption{Solutions with a space-like vector $\xi$.}
\label{tab:Sol2}
\setlength{\tabcolsep}{3pt}
\renewcommand{\arraystretch}{1.8}
\begin{tabular}{|c|l|l|l|l|l|l|}
\hline
No. & $\phi$ & $m^2$ & $\xi\cdot\xi$ & b& $D$ & $\phi_v$\\ \hline
1 & $\frac{b(X\cdot\xi)}{1-(X\cdot\xi)^2}$& $D-2$ & $\dfrac{D-1}{3}$&
$b=\dfrac{2\sqrt{2/3}\sqrt{D-4}}{\sqrt{\lambda}}$& $D \ge 4 $ &
$0$ \\
\hline
2 & $\frac{b}{(X\cdot\xi)}$& $D-2$ & $\xi\cdot\xi>0$&
$b=\sqrt{\frac{2(\xi\cdot\xi)}{\lambda}}$& $D \ge 2 $&
$0$ \\
\hline
$3^{\pm}$& $\phi_v+\frac{b}{1-(X\cdot\xi)^2}$& $3-D$ &
$\pm\dfrac{\sqrt{(D-1)(D-3)}}{\sqrt{3}}+\frac{2(D-1)}{3}$&
\shortstack{
$b=-\dfrac{ 2(\pm\sqrt{3(D-1)}+3\sqrt{D-3})}{3\sqrt{\lambda}}$}& $D\ge3$ &
$\pm\sqrt{\dfrac{D-3}{\lambda}}$ \\
\hline
\end{tabular}
\end{table}

In the open slicing one has $(X\cdot\xi)=|\xi|\cosh t\ge|\xi|$. For case 2 this
is already enough to make the solution regular in the whole patch, while for the
non-trivial solutions of cases 1 and $3^{\pm}$ one always has
$(\xi\cdot\xi)>1$, so that the denominators $1-(X\cdot\xi)^{2}$ never vanish.
In this sense the solutions of Table~\ref{tab:Sol2} are the open-slicing
counterparts of the globally regular solutions of Table~\ref{tab:Sol}. However,
these solutions have infinite action due to the divergence of the spatial volume.

\section{Solutions in the Poincaré patch}

The embedding coordinates of the Poincaré patch are given by

\[
{\displaystyle
\begin{aligned}
X_{0}&=\sinh(t)+\frac{1}{2}x^2 e^{t},\\
X_{1}&=\cosh(t)-\frac{1}{2}x^2 e^{t},\\
X_{i}&=e^{t} x_i,
\end{aligned}}
\]
and the metric takes the form
\begin{align}
    ds^2 =-dt^2+e^{2 t} d x^2.
\end{align}
Now let us write the solutions from Table~\ref{tab:Sol} that can be constructed
in the $D=4$ case, i.e. the case 2 and 4:
\begin{align}
 \phi_2=\frac{1}{\sqrt{\lambda}}
 \frac{\sinh(t)+\frac{1}{2}x^2 e^{t}}
 {\sqrt{1+\left(\sinh(t)+\frac{1}{2}x^2 e^{t}\right)^2}},
 \qquad \text{and} \quad
 \phi_4=\frac{1}{\sqrt{\lambda}}
 \left(1-\frac{4}{1+3\left(\sinh(t)+\frac{1}{2}x^2 e^{t}\right)^2}\right).
\end{align}

At the same time, one can trace the propagation in the position of the point at which $\phi=0$
for the first solution, and of the extremum for the second solution. Then it
becomes clear that this is a contracting bubble whose physical radius is initially
equal to the Hubble radius and then contracts according to
\begin{align}
R_f=\sqrt{1-e^{2 t}}.
\end{align}
At $t=0$ the points we are following reach the origin, and then the solutions
begin to flatten out for $t>0$.

\begin{figure}[H]
    \centering
    \includegraphics[scale=0.25]{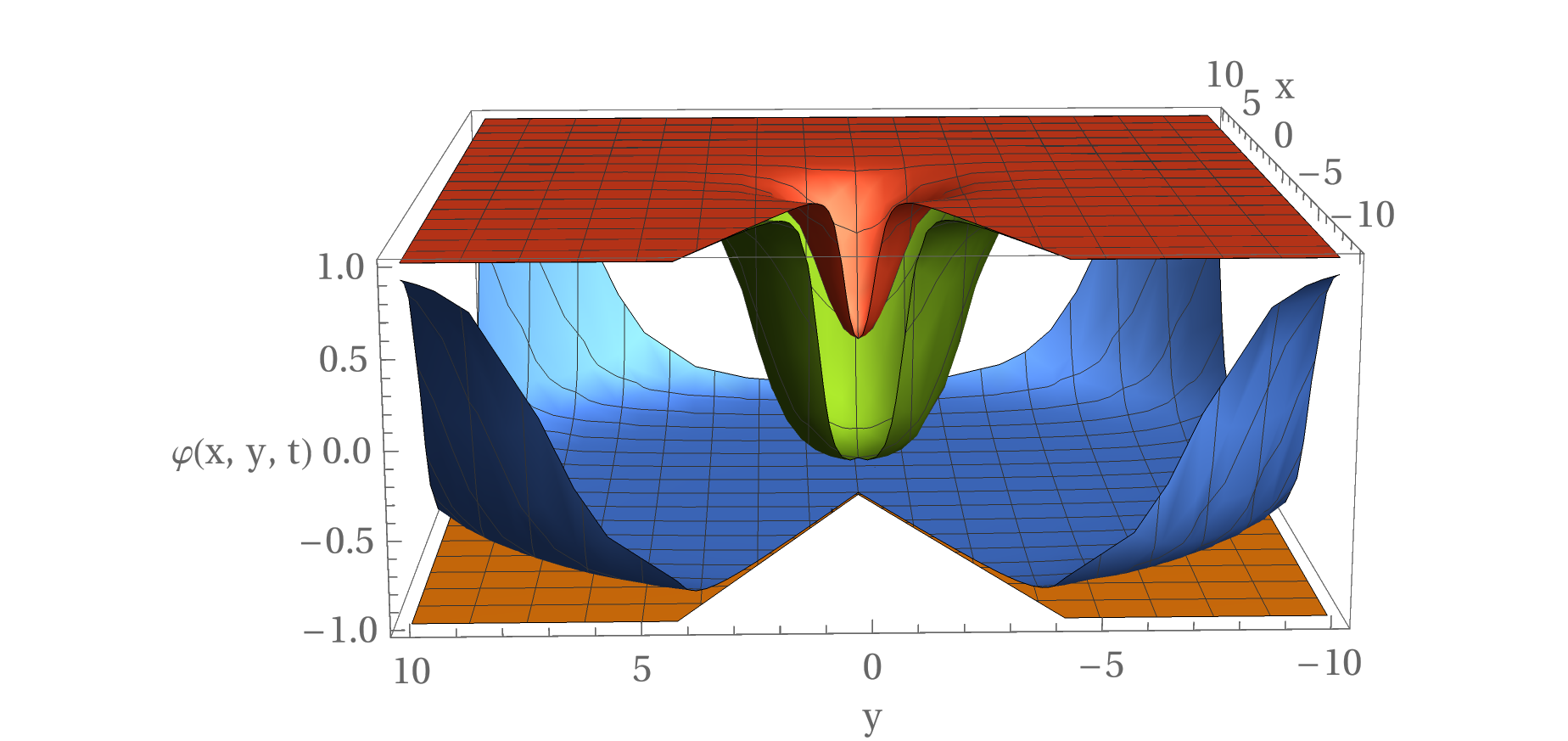},
    \includegraphics[scale=0.25]{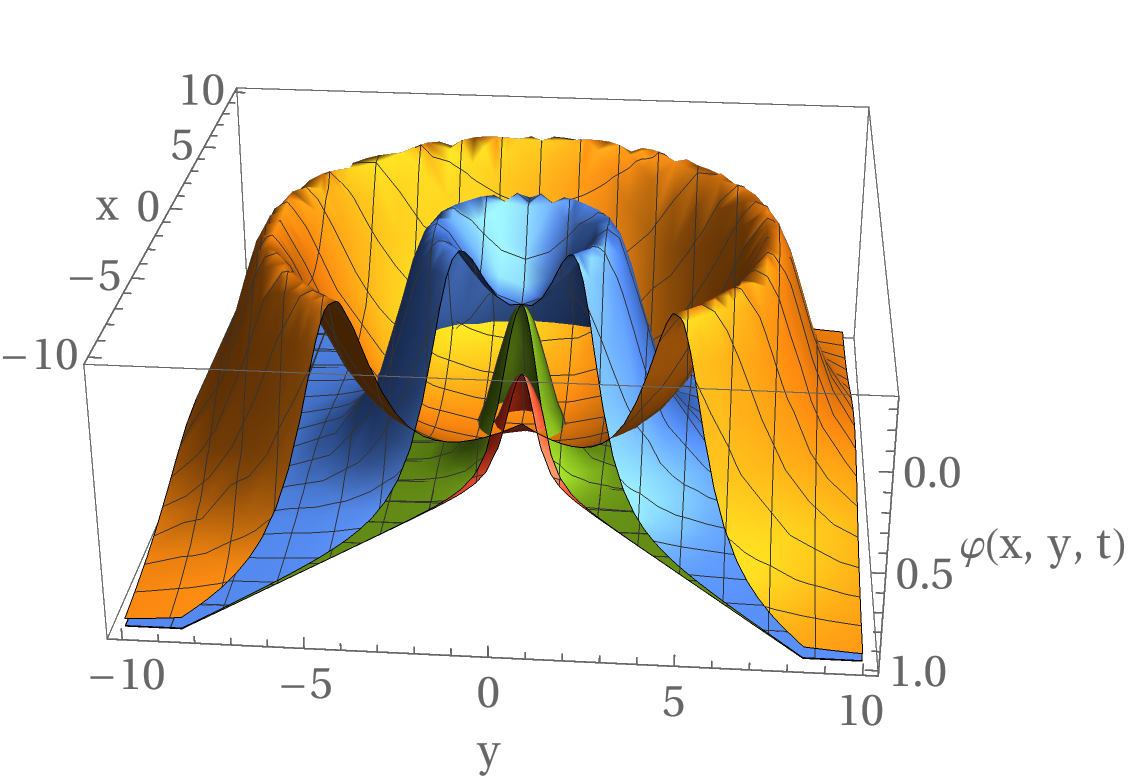}
    \caption{Profiles of the two solutions in the Poincaré region at different
    times. Orange: $t=-3$; blue: $t=-1$; green: $t=1$; red: $t=3$.}
    \label{PicOfSolinOincare}
\end{figure}

As a result, these solutions look just like spherically symmetric waves that
propagate from infinity to the origin. The speed of the extremum can be greater
than the speed of light. This is related to the fact that the propagating surface
is space-like; that is, from the point of view of the global patch it is simply a
spatial section, whereas from the point of view of Poincaré coordinates it is a
bubble with a space-like surface. The time evolution of the solutions is shown in
Figure~\ref{PicOfSolinOincare}.

\section{Stress-energy tensor}

For the solution discribing global vacuum transition we have constructed the energy-momentum tensor;
for the other solutions the answer is not as elegant. We consider the solution
\begin{align}
\phi=\sqrt{\frac{D-3}{\lambda}}\sin T
\end{align}
in the double-well potential. The energy--momentum tensor is diagonal:
\begin{align}
  T^{\mu}{}_{\nu}=\frac{D-3}{4\lambda}
  \begin{pmatrix}
    -(D-1) & 0 \\
    0 & (5-D) \mathbf{1}_{D-1}
  \end{pmatrix} \cos^4T,
\end{align}
and has the form of the stress-energy tensor of a perfect fluid with a
time-independent equation of state
\begin{align}
  w=\frac{p}{\varepsilon}=\frac{5-D}{D-1}.
\end{align}
The most interesting point is that in the $3+1$-dimensional case $(D=4)$
this looks exactly like the stress-energy tensor of radiation:
\begin{align}
  T^{\mu}_{\nu}=\frac{\cos^4T}{4\lambda} diag(-3,1,1,1),\qquad
  T^{\mu}_{\mu}=0,\qquad w=\frac13.
\end{align}

\section{Linear fluctuations on top of solutions}

In cosmology the process of particle creation may play an important role in the
mechanism of reheating after inflation \cite{Kofman:1994rk,Kofman:1997yn}, which
is believed to be responsible for the creation of almost all the particles in our
Universe. In general, particle creation occurs on time-dependent backgrounds even
for fields without self-interactions. The probability of particle creation can be
expressed in terms of the imaginary part of the effective action
\cite{Candelas:1975du,Polyakov:2007mm,Akhmedov:2009ta,Akhmedov:2019esv,Akhmedov:2024qvi,Akhmedov:2024axn}.
At tree-level particle creation in de Sitter space can be observed in even spacetime dimensions
$D = 2 N$ for massive theory. But for massless and conformally coupled case particle creation does not occure. We want ti check how the situation changes in the presence of solitons which we consider.

In conformal coordinates the metric has the following form:
\begin{align}
ds^2=\frac{1}{\cos^2 T}\left(-dT^2+d\Omega_{D-1}^2\right),
\end{align}
and the d'Alembert operator is
\begin{align}
\Box_g
=
\cos^2 T
\left(
-\partial_T^2-(D-2)\tan T \partial_T
+\Delta_{S^{D-1}}\right).
\end{align}
Now let us consider a perturbation of the solution $\phi=\phi_0+\delta \phi$ of
the equation
\begin{align}
    \Box \phi -m^2 \phi -\lambda \phi ^3=0,
\end{align}
which gives
\begin{align}
\left(\Box-m_{\rm eff}^2(x)\right)\delta\phi=0,
\qquad
m_{\rm eff}^2(x)=m^2+3\lambda\phi_0(x)^2.
\end{align}

For case 2 from Table~\ref{tab:Sol},
\begin{align}
\label{kinksol}
  \phi_0=\sqrt{\frac{D-3}{\lambda}}\sin T,
\end{align}
it is a global vacuum transition solution that interpolates between the two vacua.
For this case the effective mass is
\begin{align}
    m_{\rm eff}^2(x)= 2(D-3)-3(D-3) \cos^2(T).
\end{align}
Hence the equation of motion takes the form
\begin{align}
    \cos^2 T
\left(
-\partial_T^2-(D-2)\tan T \partial_T
+\Delta_{S^{D-1}}\right) \delta \phi
-\left( 2(D-3)-3(D-3) \cos^2(T)\right)\delta \phi=0.
\end{align}
To solve this, we expand the perturbation in the following form:
\begin{align}
\delta \phi=(\cos T)^{(D-2)/2}f_l(T) Y_{lm},
\end{align}
where $Y_{lm}$ are hyperspherical harmonics satisfying
\begin{equation}
\Delta_{S^{D-1}} Y_{lm} = -l(l+D-2) Y_{lm}.
\end{equation}
As a result, we can rewrite the equation as
\begin{align}
\label{peret}
-f_l''+\frac{\nu(\nu+1)}{\cos^2 T} f_l=E_l^2 f_l,
\end{align}
where
\begin{align}
\nu=\frac{D-6}{2},
\qquad
E_l^2=\left(l+\frac{D-2}{2}\right)^2-3(D-3).
\end{align}

At the same time, for the case
\begin{align}
\label{vacuumca}
    \phi_0 =\phi_v=\sqrt{\frac{D-3}{\lambda}},
\end{align}
the effective mass is given by
\begin{align}
    m_{\rm eff}^2(x)= 2(D-3).
\end{align}
In this case we have the same equation for the perturbation \eqref{peret}  but with a
different $E_l^2$:
\begin{align}
    E_l^2=\left(l+\frac{D-2}{2}\right)^2.
\end{align}
Hence the global vacuum transition solution simply changes the spectrum by $-3(D-3)$ for each mode.

Note also that equation \eqref{peret} is reflectionless in the case $\nu\in \mathbb{N}$, i.e.\
\begin{align}
    D=2,3,4,\ldots
\end{align}
The cases $D=4,6$ are clear since then the potential vanishes, i.e.\
$\nu(\nu+1)=0$. Note that in these cases the asymptotic effective mass
\begin{align}
    m_{\rm eff}^2(x)= 2(D-3)
\end{align}
coincides with the conformal effective mass:
\begin{align}
    m^2_{\rm eff\ conf}=\xi_c R=\frac{D(D-2)}{4}.
\end{align}

As an illustrative example, let us compute the modes explicitly for $D=4$ for
the solution \eqref{kinksol}. If $l \ge 1$:
\begin{align}
    f_l(T)=c_1 e^{-i E_l T}
    +c_2 e^{i E_l T},
\end{align}
while if  $l=0$, $E_0^2=-2$:
\begin{align}
 f_0(T)= c_1 e^{-\sqrt{2} T} +c_2 e^{\sqrt{2} T}.
\end{align}
As one can see, the $l=0$ mode does not grow
unboundedly because $T\in(-\frac{\pi}{2},\frac{\pi}{2})$. Moreover, for $l\ge 0$
we can always choose a positive-frequency mode, and therefore we have a good
choice of positive-frequency modes both at past and future infinity.

For the vacuum solution \eqref{vacuumca}, all modes are oscillating since $E_l^2>0$.

\section{Quantization}

In both cases discussed in the previous section we consider eqution of motion for pertrubations:
\begin{align}
\left(\Box-m_{\rm eff}^2(x)\right)\delta\phi=0,
\qquad
m_{\rm eff}^2(x)=m^2+3\lambda\phi_0(x)^2,
\end{align}
with the ansats: 
\begin{align}
\delta \phi=\cos( T )f_l(T) Y_{lm}.
\end{align}
Hence we can construct the field operator:
\begin{align}
    \hat{\varphi}= \cos( T ) \sum_{l,m}  f_l(T) Y_{lm}  \hat{a}_{lm}+h.c.
\end{align}
and this operator should obey the commutation relation:
\begin{align}
    \left[ \hat{\varphi}, \dot{\hat{\varphi}}\right] =i\frac{\delta(x-y) }{\sqrt{-g}g^{00}}.
\end{align}
Using that the modes should be normalized such that
\begin{align}
  \left(f_l(T) \dot{f}_l^*(T)-\dot{f}_l(T) f_l^*(T)\right) = i,
\end{align}
hence for the case $E_l^2>0$ we can choose the solution
\begin{align}
    f_l(T)= \frac{1}{\sqrt{2 E_l} }e^{-i E_l T},
\end{align}
and for the case $E_l^2<0$, i.e. for $l=0$, $E_0^2=-2$, we can choose the solution
\begin{align}
    f_0(T)=\frac{1}{\sqrt{2 |E_0|}} \left(\cosh(|E_0| T)-i \sinh(|E_0| T) \right),
\end{align}
where $|E_0|=\sqrt{2}$.

As a result we can construct the two-point function:
\begin{align}
    W(1,2)= \cos( T_1 ) \cos(T_2)\sum_{l,m}  f_l(T_1) f_l^*(T_2) Y_{lm}(\Omega_1)Y_{lm}(\Omega_2).
\end{align}

For the vacuum case $E_l=l+1$ and:
\begin{align}
    W_v(1,2)= \cos( T_1 )\cos( T_2 )\sum_{l,m}   \frac{1}{2 (l+1) }e^{-i (l+1) (T_1-T_2)} Y_{lm}(\Omega_1)Y_{lm}(\Omega_2);
\end{align}
then we can take the sum over $m$:
\begin{align}
\sum _m Y_{lm}(\Omega_1) Y_{lm} (\Omega_2) =\frac{ (l+1)}{2\pi^2}
\frac{\sin\big((l+1)\theta\big)}{\sin\theta},
\end{align}
where $\theta$ is the angle between two points on the sphere with coordinates
$\Omega_1$ and $\Omega_2$. The sum over $l$ gives the de Sitter invariant
two-point function:
\begin{align}
  \label{dsinvtwopoin}
    W_v=\frac{1}{8 \pi^2} \frac{1}{1-(X_1\cdot X_2)},
\end{align}
where
\begin{align}
    (X_1 \cdot X_2)=\frac{1}{\cos(T_1)\cos(T_2)}\left(\cos(\theta)-\sin(T_1)\sin(T_2)\right).
\end{align}
In this case one obtains the standart retarded Green's function.
At the same time for the global vacuum transition solution \eqref{kinksol}, we obtain:
\begin{gather}
    W_g= \nonumber\\
   \frac{  \cos( T_1 )\cos( T_2 )}{4 \sqrt{2}\pi^2}
   \left(\cosh(\sqrt{2} T_1)-i \sinh(\sqrt{2} T_1) \right)
   \left(\cosh(\sqrt{2} T_2)+i \sinh(\sqrt{2} T_2) \right)
    + \nonumber\\
    \frac{\cos( T_1 )\cos( T_2 ) }{2 \pi^2}\sum_{l>0}
    \frac{(l+1)}{2 \sqrt{\left(l+1\right)^2-3} }
    e^{-i \sqrt{\left(l+1\right)^2-3} (T_1-T_2)}
    \frac{\sin\big((l+1)\theta\big)}{\sin\theta}.
\end{gather}
We cannot compute this sum explicitly. But we can compute the commutator of the
field operator, which is state independent. It can be written in a compact form:
\begin{gather}
[\hat\varphi_1,\hat\varphi_2]_g=
    i\frac{\cos( T_1 )\cos( T_2 ) }{2 \pi^2}\sum_{l=0}
    \frac{(l+1)}{ \sqrt{\left(l+1\right)^2-3} }
    \sin\left( \sqrt{\left(l+1\right)^2-3} (T_2-T_1)\right)
    \frac{\sin\big((l+1)\theta\big)}{\sin\theta},
\end{gather}
or
\begin{gather}
[\hat\varphi_1,\hat\varphi_2]_g= \nonumber\\
    -i\frac{\cos( T_1 )\cos( T_2 ) }{2 \pi^2}\partial_\theta \sum_{l=0}
    \frac{1}{ \sqrt{\left(l+1\right)^2-3} }
    \sin\left( \sqrt{\left(l+1\right)^2-3} (T_2-T_1)\right)
    \frac{\cos\big((l+1)\theta\big)}{\sin\theta}.
\end{gather}
To take this sum we can use the Poisson summation formula:
\begin{align}
    \sum_{n\in\mathbb Z}g(n)e^{in\theta}=\sum_{m\in\mathbb Z}\hat g(\theta+2\pi m),
\end{align}
where $\hat g$ is the Fourier transform:
\begin{align}
    \hat g(x)=\int_{-\infty}^{\infty}dk\, g(k) e^{ikx}.
\end{align}
Using the table integral \cite{gradshteyn2007}:
\begin{equation}
\int_0^\infty\frac{\sin\big(p\sqrt{k^2+a^2}\big)}{\sqrt{k^2+a^2}}\cos(bk) dk=
\begin{cases}\frac{\pi}{2}J_0\big(a\sqrt{p^2-b^2}\big), & 0<b<p,\\ 0, & b>p,\end{cases}
\end{equation}
as a result we obtain:
\begin{align}
[\hat\varphi_1,\hat\varphi_2]_g
=i\frac{\cos T_1\cos T_2}{4\pi\sin\theta} sign(t)
\left[\delta(\theta-|t|)+\sqrt3 \theta \Theta(|t|-\theta)
\frac{I_1\big(\sqrt3\sqrt{t^2-\theta^2}\big)}{\sqrt{t^2-\theta^2}}\right],
\end{align}
where $t=T_2-T_1$. The first term is just the commutator of two fields in the
vacuum, i.e.\ it is the imaginary part of the de Sitter invariant two-point
function \eqref{dsinvtwopoin}. Hence:
\begin{align}
[\hat\varphi_1,\hat\varphi_2]_g-[\hat\varphi_1,\hat\varphi_2]_v
=i sign(t)
\Theta(|t|-\theta)\frac{\cos T_1\cos T_2}{4\pi} \sqrt3\frac{\theta}{\sin\theta}
\frac{I_1\big(\sqrt3\sqrt{t^2-\theta^2}\big)}{\sqrt{t^2-\theta^2}}.
\end{align}
Usually, in a massive theory an oscillating tail decays with time,
but here, because of the growing mode with $l=0$, the tail also grows with time. As a result, the global vacuum transition solution amplifies the signal.

\section{Conclusion}

We have obtained several families of analytic classical solutions of
$\lambda\phi^4$ theory in de Sitter space, classified them by the causal
character of $\xi$, and computed their classical actions and stress-energy
tensor. In particular, the $3+1$-dimensional global vacuum transition solution has the
radiation equation of state $w=1/3$. We show that for global vacuum transition the retarded Green's function acquires a growing tail compared to the vacuum case, implying that the solution amplifies the signal at late times.

We would like to thank E. Akhmedov and D. Sadekov for valuable discussions. The
work of Dmitrii Diakonov was supported by the grant No. 26-12-00330 from the
Russian Science Foundation (RSF).

\newpage
\bibliographystyle{unsrturl}
\bibliography{bibliography}

\end{document}